\documentclass[twoside]{article}
\usepackage{multicol}
\def\etal{{\it et al.}}
\def\eg{{\it e.g.~}}
\def\PR#1#2#3{Phys. Rev. {\bf #1}, #2 (#3)}

\def\PTP#1#2#3{Prog. Theor. Phys. {\bf #1}, #2 (#3)}

\def\vev#1{\langle #1 \rangle}
\def\gsim{~{\rlap{\lower 3.5pt\hbox{$\mathchar\sim$}}\raise 1pt\hbox{$>$}}\,}
\def\lsim{~{\rlap{\lower 3.5pt\hbox{$\mathchar\sim$}}\raise 1pt\hbox{$<$}}\,}
\def\ov{\overline}

\def\schz#1{\sin^22\theta_{_{\rm CHOOZ}}^{#1}}
\def\matm#1{\delta m^{2~#1}_{_{\rm 13}}}

\def\dmns#1{\delta_{_{\rm MNS}}^{#1}}
\def\NBBn#1{{${\rm NBB}$($#1$GeV)}}

\begin{document}
\title{
Measuring the CP-Violating Phase by a Long Baseline Neutrino Experiment
with Hyper-Kamiokande
\footnote{{%
This talk is based on Ref.\cite{H2H}.
}}
}
\author{
N.~Okamura\thanks{e-mail: nokamura@vt.edu}\\\\
{\small \it
IPPAP, Physics Department, Virginia Tech. Blacksburg, VA 24061, USA}
}
\date{\empty}
\maketitle
\vspace*{-40ex}
\begin{flushright}
\begin{tabular}{l}
{VPI-IPPAP-02-10}\\
{hep-ph/0209123}\\
\end{tabular}
\end{flushright}
\vspace*{40ex}
\vspace*{-15ex}

\begin{multicols}{2}{
We study the sensitivity of a long-base-line (LBL) experiment
with neutrino beams from the High Intensity Proton Accelerator
(HIPA) \cite{HIPA_web},
and a proposed 1Mt water-$\check {\rm C}$erenkov detector,
Hyper-Kamiokande (HK) \cite{HK}, 295km away from the HIPA,
to the CP phase ($\dmns{}$) of
the three-flavor lepton mixing
(Maki-Nakagawa-Sakata (MNS)) matrix
\cite{MNS}.
Neutrino oscillations depend on the three mixing angles,
and two mass-squared differences.
Two of the mixing angles and the mass-squared differences 
are constrained by solar and atmospheric
neutrino observations. 
For the third mixing angle
only an upper bound, $\schz{} < 0.1$,
is obtained from
the reactor neutrino experiments.

We examine a combination of the $\nu_\mu$ narrow-band beam (NBB)
at two different energies, $\vev{p_\pi}$=2, 3GeV, and
the $\ov\nu_\mu$ NBB at $\vev{p_\pi}=2$GeV.
By allocating 1Mton$\cdot$year each for the two $\nu_\mu^{}$
beams and 4Mton$\cdot$years for the $\ov\nu_\mu^{}$ beam, 
we can efficiently measure
the $\nu_\mu^{} \to \nu_e^{}$ and $\ov\nu_\mu^{} \to \ov\nu_e^{}$
transition probabilities, as well as the $\nu_\mu^{}$ and 
$\ov \nu_\mu^{}$ survival probabilities.

CP violation in the lepton sector can be established at 
the 4$\sigma$ (3$\sigma$) level if the MSW \cite{MSW}
large-mixing-angle scenario
of the solar-neutrino deficit is realized,
$|\dmns{}|$ or
$|\dmns{}-180^{\circ}| >$ 30$^{\circ}$, and 
$\schz{} > 0.03$ (0.01).
The phase $\dmns{}$ is more difficult to constrain by this experiment
if there is little CP violation, $\dmns{}\sim 0^\circ$
or $180^\circ$.
The two cases can only be distinguished
at the 1$\sigma$ level if $\schz{} \gsim 0.01$.
If we remove the \NBBn3 data from the fit, 
they cannot be distinguished even at the 1$\sigma$ level.
This two-fold ambiguity between $\dmns{}$ and $180^\circ-\dmns{}$
is found in general for all $\dmns{}$.
This two-fold ambiguity between $\dmns{}$ and $180^\circ-\dmns{}$
is found in general for all $\dmns{}$,
because the difference in the predictions can
be adjusted by a shift in the fitted $\schz{}$ value.

We point out that the low-energy LBL experiment
like HIPA-to-HK cannot determine the sign of larger mass-squared
differences, $\matm{}$,
because of the small matter effect at low energies.
If we repeat the analysis by using the same input data but 
assuming a negative sign for $\matm{}$,
we obtain another excellent fit with only a slight
shift in the model parameters.
Very long-base-line experiments, $L>1000$km,
at higher energies \cite{H2B}
are needed to determine the sign of $\matm{}$.

\noindent
{\it Acknowledgments}\\
The work of NO is supported in part by a grand from 
the US Department of Energy, DE-FG05-92ER40709.

}
\end{multicols}
\end{document}